\documentclass[printer, letter]{aa}
\usepackage{lineno}
\usepackage{lscape}
\usepackage{lipsum}
\usepackage[varg]{txfonts}
\usepackage{graphicx}
\usepackage[colorlinks=true, citecolor=blue, urlcolor=blue]{hyperref}
\usepackage{placeins}
\usepackage{color, soul}

\begin{document} 

   \title{Ultrahigh-energy cosmic-ray signature in GRB 221009A}


   \author{Saikat Das
          \inst{1}
          \and
          Soebur Razzaque\inst{2,3,4}
          }

   \institute{Center for Gravitational Physics and Quantum Information, Yukawa Institute for Theoretical Physics, Kyoto University, Kyoto 606-8502, Japan\\
              \email{saikat.das@yukawa.kyoto-u.ac.jp}
         \and
             Centre for Astro-Particle Physics (CAPP) and Department of Physics, University of Johannesburg, PO Box 524, Auckland Park 2006, South Africa\\
             \email{srazzaque@uj.ac.za}
         \and
         	 Department of Physics, The George Washington University, Washington, DC 20052, USA
         \and
             National Institute for Theoretical and Computational Sciences (NITheCS), Private Bag X1, Matieland, South Africa
             }

   \date{Received YYYY; accepted ZZZZ}

 
  \abstract
   {
   The brightest long gamma-ray burst detected so far by the Swift-BAT and Fermi-GBM telescopes, GRB~221009A, provides an unprecedented opportunity for understanding the high-energy processes in extreme transient phenomena. We find that the conventional leptonic models, synchrotron and synchrotron-self-Compton, for the afterglow emission from this source have difficulties explaining the observation of $\gtrsim 10$ TeV $\gamma$ rays, by the LHAASO detector, and extending up to 18 TeV energies. We model $\gamma$-ray spectrum estimated in the energy range 0.1-1 GeV by the Fermi-LAT detector. The flux predicted by our leptonic models is severely attenuated at $> 1$ TeV due to $\gamma\gamma$ pair production with extragalactic background light, and hence an additional component is required at $\gtrsim 10$ TeV. Ultrahigh-energy cosmic rays can be accelerated in the GRB blastwave, and their propagation induces an electromagnetic cascade in the extragalactic medium. The line of sight component of this flux can explain the emission at $\gtrsim 10$ TeV detected by LHAASO, requiring a fraction of the GRB blastwave energy in ultrahigh-energy cosmic rays. This could be an indication of ultrahigh-energy cosmic-ray acceleration in GRBs.
}

   \maketitle
%

\section{Introduction}

The origin of ultrahigh-energy cosmic rays (UHECRs; $E\gtrsim10^{17}$ eV) is an outstanding problem in physics and astrophysics. Gamma-ray bursts (GRBs), by being the most powerful electromagnetic explosions in the universe, are thought to be prime candidates to accelerate particles to ultrahigh energies \citep{Waxman:1995vg, Vietri:1995hs}. Direct signatures of UHECRs from GRBs, however, are absent. Non-detection of high-energy neutrinos from GRBs by IceCube \citep{ICGRB2015ApJ, ICGRB2017ApJ} has put severe constraints on the cosmic-ray acceleration during the prompt emission phase \citep[see, e.g.,][]{Waxman:1997ti, Razzaque2004PhRvD, Murase2006PhRvD, Zhang:2012qy}. Neutrino signature of the UHECR acceleration in the afterglow phase of GRBs is difficult to detect with the current generation of the neutrino detectors \citep{Razzaque:2013dsa, Tamborra2015JCAP, Thomas2017PhRvD}. Gamma-ray signatures from very nearby GRBs on the other hand can be used to study UHECR acceleration in these ultra-relativistic jets.

GRB~221009A is the brightest long GRB detected by the Swift Burst Alert Telescope (BAT) \citep{Dichiara_2022} and the Fermi Gamma-ray Burst Monitor (GBM) on October 9, 2022 at 13:16:59.99 UT \citep{Veres_2022}. Subsequently, the Fermi Large Area Telescope (LAT) detected $>100$~MeV $\gamma$ rays in the 200-800 s time interval after the GBM trigger ($T_0$). The highest-energy photon being of energy 99.3 GeV detected at $T_0 + 240$~s \citep{Bissaldi_2022, Pillera_2022}. This is the most energetic photon detected by Fermi-LAT from a GRB. The Large High Altitude Air Shower Observatory (LHAASO) detected over 5000 photons from GRB~221009A within $T_0 + 2000$~s in the 0.5--18~TeV range, making GRB~221009A the first GRB detected above 10 TeV \citep{Huang_2022}. At $T_0 + 4536$~s Carpet-2, a ground-based Cherenkov detector reported the detection of a 251~TeV photon from the direction of the burst \citep{dzhappuev22}. The detection of such energetic photons from even a nearby GRB such as GRB~221009A at $z=0.15$ \citep{Ugarte-Postigo_2022} is extremely interesting, given that the opacity of the universe for $\gamma$-ray propagation is very large due to $e^\pm$ pair-production with the optical/UV/IR photons of extragalactic background light (EBL) \citep{finke10_EBLmodel, Gilmore:2011ks, dominguez11}, which led to the speculation of Lorentz-invariance violation \citep{dzhappuev22, baktash22, li22, Finke2022arXiv221011261F} or mixing with axion-like particles \citep{galanti22, baktash22, troitsky22} for VHE $\gamma$ rays to evade $e^+e^-$ pair-production. 

The afterglow of GRB~221009A has also been detected by a number of X-ray telescopes, such as Swift-XRT, INTEGRAL, STIX on Solar Orbiter, IXPE, NICER, NuSTAR, etc.\ and by numerous optical telescopes around the world, and also by radio telescopes such as VLA, MeerKAT, ATCA, etc. Given the power-law nature of the  Fermi-LAT photon flux $(6.2\pm 0.4)\ \times 10^{-3}\, {\rm ph}\, {\rm cm}^{-2}\, {\rm s}^{-1}$ with a photon index of $ -1.87\pm0.04$ in the 200--800~s time window and that the LAT emission extended for about 25~ks post-GBM trigger \citep{Pillera_2022}, it is likely that $\gamma$ rays detected by LAT also originated from the afterglow. While synchrotron and synchrotron-self-Compton (SSC) processes can usually explain radio to very-high energy (VHE, $\gtrsim 100$~GeV) $\gamma$-ray observations \citep{Joshi:2019opd}, a large flux of TeV $\gamma$ rays detected by LHAASO must originate from a different mechanism. Hadronic emission mechanisms, such as proton-synchrotron radiation \citep{Razzaque2010OAJ, Wang2009ApJ, Razzaque2010ApJ, Zhang2022arXiv221105754Z} or photohadronic interactions \citep{Asano2014ApJ, Sahu2020ApJ} can produce VHE emission from the GRB, but their flux on Earth would be severely attenuated in the EBL as well.

In this work, we propose that VHE $\gamma$-rays detected by LHAASO with energy up to a few TeV is produced by SSC emission, and $\gamma$ rays above this energy are produced by UHECRs, accelerated in the GRB blastwave \citep{Waxman2000ApJ, Dai2001ApJ}. They  propagate along our line of sight and interact with the EBL and cosmic microwave background (CMB) photons to produce VHE $\gamma$ rays, in addition to the synchrotron-SSC emission. A similar method is also sometimes adopted to explain the unattenuated hard TeV spectrum of blazars \citep{Essey_10a}. The cosmogenic flux, however, is not as severely attenuated as the other components coming directly from the GRB.

\section{\label{sec:model}Gamma-ray emission models}

\subsection{\label{sec:synchro-SSC}Synchro-Compton emission}

The total isotropic $\gamma$-ray energy of GRB~221009A has been estimated to be $(2-6)\times 10^{54}$~erg \citep{Ugarte-Postigo_2022, Kann_2022}. Therefore, for the afterglow emission from GRB~221009A, we use an adiabatic blastwave with kinetic energy $E_k = 10^{55}E_{55}$~erg evolving in a constant density interstellar environment \citep{blandford_mckee_soln_76}. We calculate the synchrotron and SSC spectra using formulas in \cite{Joshi:2019opd}, which are based on the models in Refs.~\cite{sync_SPN_aftrglw, sari_IC_paper}. For the time-dependent synchrotron spectrum, relevant break energies are that from the electrons of minimum Lorentz factor, cooling Lorentz factor, and saturation Lorentz factors. For modeling the 0.1--1 GeV $\gamma$-ray flux from Fermi-LAT, these energies are given by
\begin{align}
    E_m &= 28.6\, \epsilon_{e,-1.5}^2 \epsilon_{B,-1.8}^{1/2} E_{55}^{1/2} t_{2.7}^{-3/2} ~{\rm eV}
    \nonumber \\
    E_c &= 3.9\, \epsilon_{B,-1.8}^{-3/2} E_{55}^{-1/2} n_{-3.7}^{-1} t_{2.7}^{-1/2} ~{\rm keV}
    \nonumber \\
    E_s &= 4.6\, \phi^{-1} E_{55}^{1/8} n_{-3.7}^{-1/8} t_{2.7}^{-3/8} ~{\rm GeV} \, ,
    \label{eq:syn_breaks}
\end{align}
at $t=10^{2.7}t_{2.7}$~s post-trigger, when the blastwave is in a decelerating phase \citep{blandford_mckee_soln_76}. Here we have assumed the fraction of the shock energy in non-thermal electrons as $\epsilon_e = 10^{-1.5}\epsilon_{e,-1.5}$ and in a turbulent magnetic field as $\epsilon_B = 10^{-1.8}\epsilon_{B,-1.8}$. The Compton parameter $Y \approx \sqrt{\epsilon_e/\epsilon_B} = 1.4$ in our modeling for a slow-cooling ($E_m < E_c$) synchrotron spectrum. The electrons follow a power-law distribution of Lorentz factor $\gamma^{-p}$, where we have assumed $p = 1.74$. We have also assumed the interstellar medium has a rather low particle density $n = 10^{-3.7}n_{-3.7}$~cm$^{-3}$. We included SSC cooling while calculating $E_c$ and an efficiency factor $\phi^{-1} \lesssim 1$ for electron acceleration to the maximum energy $E_s$ in Equation~(\ref{eq:syn_breaks}). Note that there is significant degeneracy among the model parameters and other sets of parameters may also produce similar fits. Our chosen set of parameters, which are within the typical range for GRB afterglows, produces the estimated Fermi-LAT flux,
\begin{align}
    E^2 \left( \frac{dN}{dE} \right) &= 1.2\times 10^{-6} \left( \frac{E}{\rm GeV} \right)^{0.13} {\rm erg}~{\rm cm}^{-2}~{\rm s}^{-1}; \nonumber \\
    &~~~~~~~~~~~~~~~~~~~~~~~~~~~~~~~~~~ E_c \le E < E_s \,,
\end{align}
in the 0.1--1 GeV range in the 200--800~s interval, post-trigger. 
%
The break energies in the SSC spectrum can be calculated with simplified assumptions as \citep{Joshi:2019opd}
\begin{align}
    E_{m, {\rm SSC}} &= 2.8\, \epsilon_{e,-1.5}^4 \epsilon_{B,-1.8}^{1/2} E_{55}^{3/4} n_{-3.7}^{-1/4} t_{2.7}^{-9/4} ~{\rm GeV}
    \nonumber \\
    E_{c, {\rm SSC}} &= 52.9\, \epsilon_{B,-1.8}^{-7/2} E_{55}^{-5/4} n_{-3.7}^{-9/4} t_{2.7}^{-1/4} ~{\rm TeV} \,.
\end{align}
The Klein-Nishina effect, however sets in at an energy
\begin{align}
    E_{\rm KN, SSC} = 1.3\, \epsilon_{B,-1.8}^{3/2} E_{55}^{3/4} n_{-3.7}^{3/4} t_{2.7}^{-1/4} ~{\rm TeV} \,,
\end{align}
and simple Thomson approximations cannot be used above this energy. Therefore, an SSC component can be estimated as below
\begin{align}
    E^2 \left( \frac{dN}{dE} \right) &= 2.0\times 10^{-6}~{\rm erg}~{\rm cm}^{-2}~{\rm s}^{-1} 
    \\ 
    & ~~~~~ \times 
    \begin{cases}
        \left( \frac{E}{E_{m,\rm SSC}} \right)^{4/3}; E \le E_{m,\rm SSC} \\
        \left( \frac{E}{E_{m,\rm SSC}} \right)^{0.63}; E_{m, {\rm SSC}} \le E \le E_{\rm KN, SSC} \,.
    \end{cases} \nonumber 
\end{align}
However, most recent EBL models predict a suppression of $\gamma$-ray flux above $\approx 100$~GeV for $z = 0.15$. The SSC flux at 18 TeV, the maximum photon energy reported by LHAASO is inadequate to explain the VHE observations. We show the synchrotron and EBL attenuated SSC fluxes in Fig.~\ref{fig:def} by dashed orange and the dashed brown lines respectively. We present the results for a higher $\epsilon_e$ value to increase the SSC flux without violating the 0.1-1 GeV flux detected by Fermi-LAT, which we model as synchrotron emission. The lower bound of the shaded region in the plot corresponds to a lower  $\epsilon_e= 10^{-2.5}\epsilon_{e,-2.5}$ and $\epsilon_B=10^{-4}\epsilon_{B, -4}$, adjusted such that the Fermi-LAT flux, modeled as synchrotron emission, is not violated. To extend the SSC flux to even higher energy by reducing the $\epsilon_B$ value considered here may also increase the synchrotron flux and thus violate the Fermi-LAT flux level. Note that the detection of a 99.3 GeV photon by Fermi-LAT at $T_0 + 240$~s is broadly consistent with the SSC flux component. 

\subsection{\label{sec:uhecr}Line-of-sight emission from UHECRs}

\begin{figure}
\centering
\includegraphics[width=0.48\textwidth]{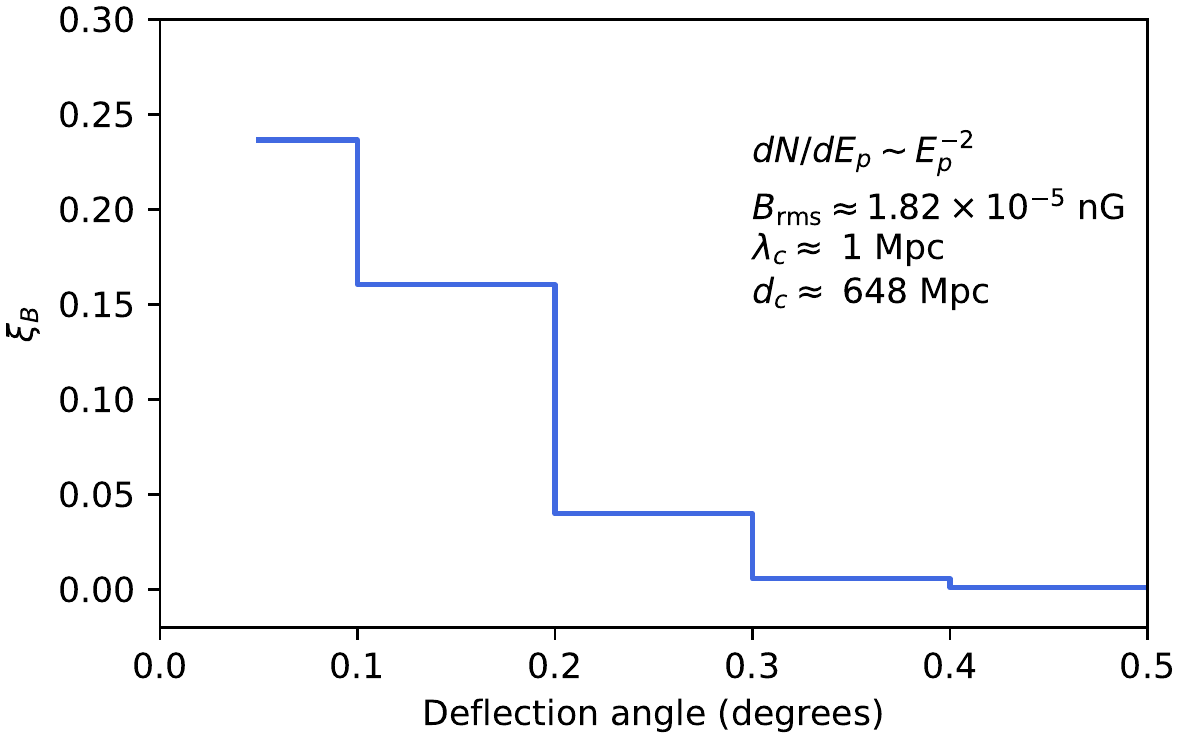}
\caption{\small{Distribution of UHECR fraction as a function of the deflection angle on the surface of a sphere centered at Earth and of radius 1 Mpc.}}
\label{fig:def}
\end{figure}

\begin{figure}
\centering
\includegraphics[width=0.48\textwidth]{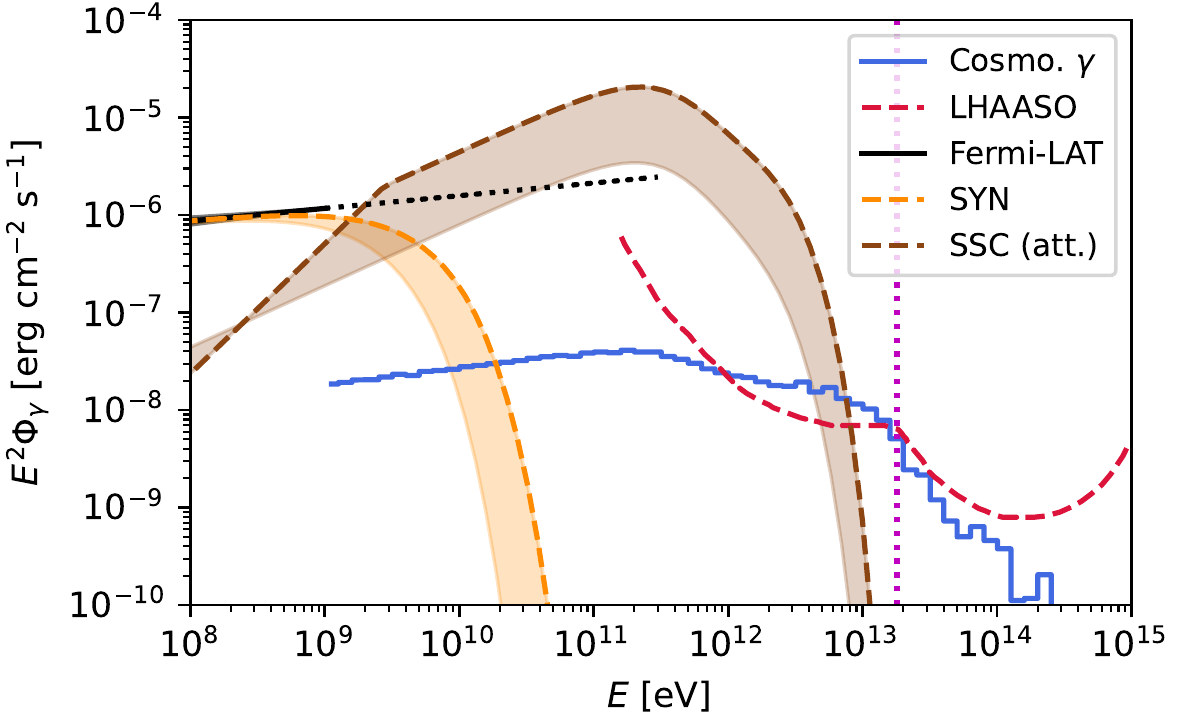}
\caption{\small{Line-of-sight cosmogenic $\gamma$-ray flux from UHECR interactions (blue curve). The black solid line corresponds to the Fermi-LAT preliminary flux estimate from GRB~221009A \citep{Pillera_2022}. The red dashed curve indicates the LHAASO sensitivity corresponding to 2000~s of observation. The dotted vertical line corresponds to the highest energy detection by LHAASO. The synchrotron and SSC emission components are shown as orange and brown dashed curves, respectively. 
}}
\label{fig:uhe}
\end{figure}

UHECRs, accelerated in the internal shocks of GRBs are expected to produce PeV neutrinos by interacting with the prompt $\gamma$ rays \citep{Waxman:1997ti}. Non-detection of these neutrinos from GRB~221009A by IceCube, within three hours around the Fermi-GBM trigger, has led to a time-integrated flux upper limit of $3.9\times 10^{-2}$~GeV/cm$^{-2}$ in the 0.8--1 PeV energy range at 90\% C.L. \citep{icecube2022GCN}, which has implications for GRB model parameters~\citep{Liu2022arXiv221114200L, Ai:2022kvd, Murase:2022vqf}. 
Here we consider UHECR acceleration in the external shock of the GRB blastwave, during the afterglow emission phase. The maximum proton energy for an adiabatic blastwave in a constant density environment can be calculated as~\citep[see, e.g.,][]{Razzaque:2013dsa}
\begin{equation}
    E = 9.7\times 10^{19}~ \phi^{-1}  \epsilon_{B,-1.8}^{1/2} E_{55}^{3/8} n_{-3.7}^{1/8} t_{2.7}^{-1/8} ~{\rm eV}.
\end{equation}
By interacting with the afterglow photons, these protons can produce neutrinos in the EeV range \citep{Waxman2000ApJ, Dai2001ApJ, Razzaque:2013dsa}. Thus the IceCube flux upper limit in the 0.8--1 PeV energy range does not apply in our scenario. 

We assume the UHECR protons accelerated in the GRB blastwave escape from the source and propagate through the extragalactic medium from their sources to Earth. Their interactions lead to the production of secondary electromagnetic (EM) particles ($e^\pm$, $\gamma$), which can initiate EM cascade undergoing various energy loss processes, such as, pair production, including double and triple pair production, inverse-Compton scattering of background photons to higher energy, etc. The extragalactic magnetic field (EGMF) can deflect the UHECRs away from our line of sight and thus the resultant flux at Earth can be a fraction of the emitted flux. The time delay induced by the deflection in EGMF can be expressed as \citep{Dermer2009NJPh},
\begin{align}
\Delta t_{\rm IGM} \approx \dfrac{d_c^3}{24r_L^2 c N_{\rm inv}^{3/2}} \approx2000\text{ s} \bigg(\dfrac{d_c}{648\text{ Mpc}}\bigg)^{3/2}  \nonumber \\
\times \bigg(\dfrac{\lambda_c}{1\text{ Mpc}}\bigg)^{3/2}
\bigg(\dfrac{B}{1.82\times10^{-5}\text{ nG}}\bigg)^2 \bigg(\dfrac{E}{100\text{ EeV}}\bigg)^2
\label{eqn:delay_igm}
\end{align}
where $d_c$ is the comoving distance of the source, which in our case is $\simeq$ 648 Mpc for the standard flat, $\Lambda$CDM cosmological parameters corresponding to a redshift $z\approx 0.151$. The number of inversions in the magnetic field $N_{\rm inv}$ is expressed as $max(d_c/\lambda_c,1)$, where $\lambda_c$ is the turbulent correlation length of the EGMF. The above expression yields the minimum time delay corresponding to the highest energy protons. The chosen parameter values thus give a time-delay consistent with the LHAASO observation time \citep{Huang_2022}. 

We use \textsc{CRPropa3.2} numerical framework for extragalactic propagation of UHECRs \citep{AlvesBatista:2016vpy, AlvesBatista:2022vem}. 
For our simulation we assume an RMS magnetic field strength of $B_{\rm rms}\approx1.82\times10^{-5}$ nG, and a coherence length of $\lambda_c \sim 1$ Mpc, so that $\Delta t \simeq 2000$s. 
To calculate the line of sight component of the EM cascade, we employ a numerical method similar to that explained in Ref.~\cite{Das:2019gtu}. We consider an observing sphere around the earth of radius 1 Mpc, i.e., the same as the coherence length, so that the deflection inside this sphere is negligible. We calculate the fraction of UHECRs that survives within $0^\circ.1$ of the initial emission direction, on the surface of this sphere. We denote this fraction as $\xi_B$. Then the line of sight component of the cosmogenic $\gamma$-ray flux  would be the fraction $\xi_B$ of the entire EM cascade arising from the UHECR propagation, obtained from a 1D simulation. We include all energy loss processes of primaries and secondary EM particles in the simulations involved with a proton spectrum of the form $dN/dE_p\sim E^{-2}$ in the energy range 0.1-100 EeV, and a random turbulent EGMF, given by a Kolmogorov power spectrum. The distribution of the UHECR fraction as a function of the deflection angle is shown in Fig.~\ref{fig:def}. We use the Gilmore et al. EBL model \citep{Gilmore:2011ks} and the Protheroe and Biermann model for the universal radio background \citep{Protheroe:1996si}.

We linearly scale the 1-yr flux sensitivity of LHAASO to Crab-like point sources \citep{Vernetto:2016gro}, as a conservative estimate to represent the GRB~221009A detection potential in 2000~s, corresponding to the time delay $\Delta t$. In absence of precise flux measurements at these energies, our presentation implies the lower limit to VHE flux from UHECR interactions. The corresponding UHECR luminosity in the energy range from 0.1-100 EeV can be presented as
\begin{equation}
L_{{\rm UHE}p} \gtrsim \dfrac{2\pi d_L^2 (1-\cos\theta_j)}{\xi_B f_{\gamma,p}} \int_{\text{1 GeV}}^{\text{100 EeV}}\epsilon_\gamma \dfrac{dn}{d\epsilon_\gamma dA dt} d\epsilon_\gamma
\label{eqn:lum}
\end{equation}
where $2\pi d_L^2 (1-\cos\theta_j)$ is the area subtended by the GRB jet at the distance of the observer. The jet opening angle is assumed to be a typical value of $6^\circ$, appropriate for GRBs \citep{Frail2001ApJ}. $f_{\gamma,p}$ is the fraction of UHECR energy going into cosmogenic $\gamma$-rays between 1 GeV and 100 EeV. The integration is over the required flux of VHE $\gamma$-rays normalized to the LHAASO sensitivity at 18 TeV. The value of $\xi_B$ within $0^\circ.1$ is found to be 0.24 and the value of $f_{\gamma,p}$ corresponding to $z=0.15$ is found to be 0.04. Using these values, we get from Equation~(\ref{eqn:lum}), $L_{{\rm UHE}p}\gtrsim 5.4\times10^{47}$ erg/s. This is the actual luminosity required in UHE protons to produce line-of-sight VHE $\gamma$-ray emission, i.e., the luminosity after the beaming correction. For $T_0+2000$~s LHASSO detection, it corresponds to an isotropic energy release of $\gtrsim 3.9\times 10^{53}$~erg in UHECR protons, a small fraction of the total kinetic energy of the blastwave.

No track-like event, having positional coincidencce with this GRB, was found by the IceCube neutrino observatory in 2 hrs from the initial trigger recorded by Fermi-GBM. IceCube has derived a time-integrated $\nu_\mu$ flux upper limit for this source at 90\% C.L., assuming a $E^{-2}$ power law \citep{IceCube_2022}. We also calculate the line-of-sight all-flavor cosmogenic neutrino flux from the GRB arriving on Earth. We use the same normalization as required for the VHE $\gamma$-rays from EM cascade and find that the neutrino fluence during 2 hrs of IceCube observation is orders of magnitude lower than the IceCube upper limit. 

%
%

\section{\label{sec:summary}Discussion and summary}

GRBs have long been considered prominent candidates for UHECR acceleration. In the blastwave model, the relativistically expanding ejecta from a central engine slows down, after prompt emission, by interacting with the ambient medium. This produces a forward shock in the decelerating blastwave, whereby protons can be accelerated to ultrahigh-energies.
The delayed high-energy $\gamma$-ray emission observed from the GRBs at $\gtrsim 1$ TeV energies can be efficiently explained by the interaction of UHECRs due to extragalactic propagation.

We find that in the case of the recent GRB~221009A, the leptonic emission due to synchrotron and SSC emission is difficult to extend up to energies as high as $\gtrsim10$ TeV. The SSC emission at the highest energies becomes inefficient due to the Klein-Nishina effect and the flux is also attenuated due to $\gamma\gamma$ pair production with the EBL photons. In our analysis, the SSC spectrum falls off sharply beyond $\sim 220$ GeV. However, the SSC spectrum is consistent with Fermi-LAT observation of $\sim 100$ GeV photon. It is noteworthy, that the SSC flux is well within the reach of LHAASO flux sensitivity normalized for 2000~s of observation. Beyond 10 TeV, any significant flux from the source is unlikely to have originated directly from the GRB blastwave, due to EBL attenuation. For this reason, we invoke the line-of-sight UHECR interactions as the origin of $\gtrsim10$ TeV $\gamma$-rays detected by LHAASO. We adjust the RMS strength of EGMF to be $B_{\rm rms} \approx 1.82\times10^{-14}$~G, such that the time delay induced by UHECR propagation from the initial trigger is comparable to $\sim2000$ s. Our estimate for the lower limit of proton luminosity is a fraction of the blastwave kinetic energy.

There can be an additional time delay of UHECRs due to propagation in the host galaxy \citep{Takami_2012}, which for GRB~221009A at $z < 1$ can be a compact galaxy \citep{Schneider2022A&A} and the magnetic field for such galaxies is unknown. For a Milky Way-type galaxy, the time delay can be expressed as \citep{Dermer2009NJPh},
%
%
$\Delta t_{\rm gal} \approx 2.25\times 10^9 (h_{\rm md}/0.1\text{ kpc})^3 (B_{\rm rms}/\mu\text{G})^2 (E/100 \text{ EeV})^2$~s, where $h_{\rm md}$ is the characteristic height of the magnetic disk and $b$ is the Galactic latitude of the UHECR source. It can be seen that for a time delay of the order of $\sim10^{3}$ s, the host galaxy's magnetic field needs to be as low as 1~nG, similar to the field in protogalaxies \citep{Beck2013}. Alternatively, for our model to be valid, the GRB needs to be positioned at the outskirts of the host galaxy or away from the disk region, so that the magnetic field is diminished. Similar assumptions are also made in other studies \citep{AlvesBatista:2022kpg}, which shows that the cascade emission induced by heavier nuclei can extend up to higher energies than LHAASO detection.

For a Bethe-Heitler dominated cascade, similar to the pair-echo effect \citep{Razzaque2004ApJ, Murase:2011cy}, the $e^\pm$ pairs produced nearer to the source can be deflected significantly, and hence induce higher time delay in the VHE $\gamma$-ray signal than that follows from Equation~(\ref{eqn:delay_igm}). However, the contribution from these pairs is less significant at $\gtrsim 10$ TeV energies, because of EBL attenuation. Thus, in our model, we assume that the production of secondaries is dominant nearer to the observer. The secondary EM particles with higher deflection, and thus higher time delay, are rejected by the line-of-sight survival fraction considered here. However, for protons with energy larger than GZK energies, there can be interaction near to the source, at $\sim100$ Mpc, and hence the actual RMS value of the EGMF needs to be lower than that estimated here. 

Observation of GRB~221009A by LHAASO at $> 10$~TeV provides a unique opportunity to probe particle acceleration and emission mechanisms of GRBs. We find for the first time UHECR acceleration signature in GRBs by explaining VHE $\gamma$-ray data.

\begin{acknowledgements}
We thank K. Murase, T.~Aldowma, E.~Burns, J.~D.~Finke, N.~Omodei, P. Reichherzer, J. D\"orner, and T. Wada for useful discussions. The work of S.D. was supported by JSPS KAKENHI Grant Number 20H05852 and by the University of Johannesburg (UJ) URC grant. Numerical computation in this work was carried out at the Yukawa Institute Computer Facility. S.R.\ was supported by grants from NITheCS and UJ URC.
\end{acknowledgements}

%
%

\bibliographystyle{aa} 
\bibliography{aa.bib}

\end{document}